\documentclass[a4paper,11pt]{article}
\pdfoutput=1 

\usepackage{jcappub} 

\usepackage[utf8]{inputenc}
\usepackage[english]{babel}
\usepackage[T1]{fontenc}
\usepackage{bbm}
\usepackage[margin=2.5cm]{geometry}               
\usepackage{amssymb}
\usepackage{amsmath}	
\newcommand\numberthis{\addtocounter{equation}{1}\tag{\theequation}}

\usepackage{graphicx}	
\usepackage{wrapfig}
\usepackage{braket}
\usepackage{stackengine,scalerel}

\usepackage[]{cleveref}

\title{\boldmath New Dynamical Degrees of Freedom from Invertible Transformations}

\author[a]{Pavel Jirou\v{s}ek,}
\author[c]{Keigo Shimada,}
\author[b]{Alexander Vikman}
\author[c]{and \qquad Masahide Yamaguchi}

\affiliation[a]{\it High Energy Physics, Cosmology \& Astrophysics Theory Group, University of Cape Town,\\Private Bag, Cape Town, South Africa, 7700}
\affiliation[b]{\it CEICO - Central European Institute for Cosmology and Fundamental Physics,\\ FZU - Institute of Physics of the Czech Academy of Sciences,\\Na Slovance 1999/2, 18220 Prague 8, Czech Republic}
\affiliation[c]{\it Department of Physics, Tokyo Institute of Technology,\\Tokyo 152-8551, Japan}

\emailAdd{pavel.jirousek@uct.ac.za}
\emailAdd{shimada.k.ah@m.titech.ac.jp}
\emailAdd{vikman@fzu.cz}
\emailAdd{gucci@phys.titech.ac.jp}

\abstract{We show that invertible transformations of dynamical variables can change the number of dynamical degrees of freedom. Moreover, even in cases when the number of dynamical degrees of freedom remains unchanged, the resulting dynamics can be essentially different from the one of the system prior to transformation. After giving concrete examples in point particle cases, we discuss changes in dynamics due to invertible disformal transformations of the metric in gravitational theories.
}

\begin{document}
\maketitle
\flushbottom
\section{Introduction}

Transformations of variables provide one of the main tools in theoretical physics. It is always advantageous to formulate a theory in such a way that it is covariant with respect to a large class of transformations. For instance, Lagrangian mechanics allows for general coordinate transformations contrary to the usual Cartesian coordinates of Newtonian mechanics, whereas the Hamiltonian canonical description extends permitted transformations even further. Among different transformations, the \emph{invertible} ones play a special role. Here we define an invertible transformation such that it has a one-to-one correspondence between the original variables and the new ones. One can expect that such one-to-one correspondence automatically implies that the dynamics of the systems before and after the transformation would be essentially (physically) the same. Hence, we could only analyze one set of variables to understand the dynamics. In fact, it is shown in Ref.~\cite{Takahashi:2017zgr} that any solution of the Euler-Lagrange equations in an original system is mapped to a solution of the Euler-Lagrange equations in a new system, and vice versa provided the transformation is (regular) invertible. 
The main conclusion of our work is that this is not always the case for invertible transformations. That is, an invertible transformation of variables or fields can change the number of dynamical degrees of freedom (DoF). Moreover, even in the case when the number of dynamical degrees of freedom is preserved, completely different dynamics can emerge after an invertible transformation.

The invertibility of a transformation can be judged by the famous inverse function theorem, which states that, as long as the Jacobian (the determinant of the Jacobian matrix) of the transformation is non-vanishing (that is, the transformation is regular), such a transformation is invertible (at least locally).\footnote{The extension of this inverse function theorem to the case of transformations with derivatives was recently given in Refs.~\cite{Babichev:2019twf,Babichev:2021bim}.} However, it should be noticed that the inverse function theorem provides only a \emph{sufficient} condition for the invertibility. In fact, even if a transformation is singular, that is, its Jacobian vanishes at some point, the transformation can be invertible.
A well-known example for such a case is a transformation $y(x) = x^{2n+1}$ with integer $n$ larger than unity, which is actually invertible even though the Jacobian of the transformation vanishes at the origin $(x = y = 0)$.

In our previous paper \cite{Jirousek:2022rym}, we have shown that the essential reason why a new DoF appears in mimetic gravity \cite{Chamseddine:2013kea} lies in the singular nature (that is, in the vanishing Jacobian) of a transformation whichever it is invertible or non-invertible. Later 
Ref.~\cite{Golovnev:2022jts} fully confirmed our results. In the current paper, we will explicitly show that \emph{singular} but \emph{invertible} transformations can actually change the number of the DoF.

The organization of this paper is as follows. In the next section, we begin with an example of a simple non-relativistic point particle system in which an invertible transformation actually changes the number of dynamical degrees of freedom. This finding is confirmed by the Hamiltonian analysis. Another interesting example is also given, where a completely new dynamics appears after an invertible transformation with keeping the number of dynamical degrees of freedom. In section \ref{sec:relativistic}, we approach closer to metric transformations and consider a polynomial time-reparametrisation invariant action for relativistic point particle and perform an \emph{invertible} transformation of the einbein. We demonstrate that, despite invertibility, this transformation gives rise to two extra DoF. In section \ref{sec:mimetic}, we provide a concrete example of how mimetic gravity \cite{Chamseddine:2013kea} can appear via an \emph{invertible} conformal transformation of the metric. Mimetic gravity is a popular framework, for review see e.g. \cite{Sebastiani:2016ras}, and in the simplest cases it is able to provide a description of a fluid-like dust \cite{Golovnev:2013jxa,Barvinsky:2013mea} useful to model dark matter. Then, in section \ref{New_K-essence} we considered the effect of the invertible field transformation on as general scalar field models as k-essence, see \cite{Armendariz-Picon:2000nqq,Armendariz-Picon:1999hyi,Chiba:1999ka}). We show that the resulting theory can avoid the Ostrogradsky instability \cite{Woodard:2006nt,Woodard:2015zca} and allows for new nontrivial vacua. The final section is devoted to conclusions and discussion. In the appendix, we also give an example of transformation with higher derivatives.

\section{Adding Degrees of Freedom to Non-Relativistic Free Particle}
\label{sec:toy}
Let us demonstrate how one can gain extra DoF on a simple ordinary mechanics example - a unit mass free particle, with the action
\begin{equation}
    S\left[q\right]=\frac{1}{2}\int_{t_{1}}^{t_{2}} dt\,\dot{q}^{2}\,,\label{eq: ex1: action}
\end{equation}
where we use standard notation $\dot{q}\equiv dq/dt$. This system clearly has only a single degree of freedom, which is reflected in the presence of exactly two integration constants in the solutions of its equation of motion $\Ddot{q}=0$. These constants are, of course, the initial velocity $v_{0}$ and the initial position $q_{0}$ in the general solution 
\begin{equation}
    q(t)=q_{0}+v_{0}\,t\ .\label{eq: ex1: classical solution}
\end{equation}
Now we introduce a non-trivial but \emph{invertible} transformation of the variable $Q\leftrightarrow q$ that will be facilitated by another variable $\phi$
\begin{equation}
    q(Q,\dot{\phi})=Q^{3}+\dot{\phi}\ .\label{eq: ex1: transformation}
\end{equation}
This transformation is clearly invertible since for any given $\phi$ we are able to find an inverse transformation
\begin{equation}
    Q(q,\dot{\phi})=\sqrt[3]{q-\dot{\phi}}\ .\label{eq: ex1: inverse transformation}
\end{equation}
The existence of this inverse is, however, not implied by the inverse function theorem. Indeed, due to the third power of $Q$ in \eqref{eq: ex1: transformation} the Jacobian of the transformation
\begin{equation}
    J=\frac{\partial q}{\partial Q}=3Q^{2}\ ,\label{eq: ex1: Jacobian}
\end{equation}
has a singular point at $Q=0$, which violates the assumptions of the inverse function theorem. Crucially, this singular point then necessarily appears as a solution of the $Q$-equation of motion. Indeed, by varying the transformed action 
\begin{equation}
    S\left[Q,\phi\right]=\frac{1}{2}\int_{t_{1}}^{t_{2}}dt\,\left(\ddot{\phi}+3Q^{2}\dot{Q}\right)^{2}\,,\label{eq: ex1: first transformed action}
\end{equation}
with respect to both $Q$ and $\phi$ we obtain the following set of equations of motion 
\begin{align}
    &\frac{\delta S}{\delta Q}=-\ddot{q}\,J=0\ ,\label{eq: ex1: y eom}\\
    &\frac{\delta S}{\delta\phi}=\frac{d}{dt}\ddot{q}=0\ .\label{eq: ex1: phi eom}
\end{align}
Clearly, the solutions of seed theory \eqref{eq: ex1: action} are also solutions of these equations of motion. However, these are not the only possible solutions. Indeed, the $Q$-equation of motion either reproduces the equation of motion of the original untransformed system ($\Ddot{q}=0$), or requires the Jacobian to vanish ($J=0$). Hence, there are \emph{two branches} of solutions: the \emph{regular branch} describing the original dynamics and the \emph{singular branch} corresponding to the singular point of the Jacobian ($Q=0$).

In the regular branch, we are away from the singular point of the invertible transformation and the dynamics are equivalent to those of the original seed theory. This is reflected in the $\phi$-equation of motion \eqref{eq: ex1: phi eom} being satisfied identically on the original solution $\ddot{q}=0$. This is rather natural, since there was only a single equation of motion prior to transformation \eqref{eq: ex1: transformation}, there is only a single non-trivial equation after transformation. Nevertheless, we are  using two variables instead of one to describe the original system - so we must ask what happens to this redundancy. The answer is that there is an obvious gauge freedom in action \eqref{eq: ex1: first transformed action}
\begin{equation}
\phi\rightarrow\phi+\epsilon\,,\quad\text{along with }\quad Q^{3}\rightarrow Q^{3}-\dot{\epsilon}\ ,\label{eq: ex1: gauge freedom}
\end{equation}
where $\epsilon(t)$ is an arbitrary function of time. This gauge freedom in action \eqref{eq: ex1: first transformed action} corresponds to the identity
\begin{equation}
\frac{\delta S}{\delta\phi}+\frac{d}{dt}\frac{\delta S}{\delta Q^{3}}=0\,.\label{eq: ex1: identity}
\end{equation}
From the perspective of theory given by action \eqref{eq: ex1: first transformed action} with two variables, the original variable $q$ given by combination \eqref{eq: ex1: transformation} is a gauge invariant variable. It is crucial that gauge transformation \eqref{eq: ex1: gauge freedom} cannot be linearized around the singular point $Q=0$. 
Moreover, it is clear that equation of motion \eqref{eq: ex1: y eom} is not gauge invariant, due to the presence of the Jacobian. Though, the equation of motion becomes gauge invariant when $Q\neq0$. 
Note that \eqref{eq: ex1: y eom} has $\Ddot{Q}$, while \eqref{eq: ex1: phi eom} has $\ddddot{\phi}$ so that naively one would count that there is at most two DoF from \eqref{eq: ex1: phi eom} and one DoF from \eqref{eq: ex1: y eom}, however one DoF can be eliminated due to gauge freedom \eqref{eq: ex1: gauge freedom} so that maximum two DoF may remain.

In the singular branch, $\phi$-equation of motion reduces to
\begin{equation}
    \frac{d^{4}}{dt^{4}}\,\phi=0\ ,
\end{equation}
which yields the solution
\begin{equation}
    \phi(t)=c_{0}+c_{1}t+c_{2}\frac{t^{2}}{2}+c_{3}\frac{t^{3}}{6} \ ,
\end{equation}
where $c_{i}$ are four constants of integration. 
A crucial point here is that transformation \eqref{eq: ex1: transformation} does not become trivial at the singular point. Instead, it reduces to
\begin{equation}
    q(Q,\dot{\phi})|_{Q=0}=\dot{\phi}\ .
\end{equation}
Thus, the gauge invariant quantity $q$ depends on three integration constants. More importantly, the motion $q(t)$ instead of a \emph{free particle} \eqref{eq: ex1: classical solution}, describes now an \emph{accelerated particle} 
\begin{equation} 
    q(t)=c_{1}+c_{2}t+c_{3}\frac{t^{2}}{2}\ ,
\end{equation}
where the constant acceleration $c_3$ should be provided by initial conditions. 
Note that the correct formulation of the action principle in new variables in \eqref{eq: ex1: first transformed action} requires standard boundary conditions corresponding to  
\begin{equation} 
    \delta Q(t_{1})=\delta Q(t_{2})=0\,,\quad\text{and}\quad\delta\phi(t_{1})=\delta\phi(t_{2})=\delta\dot{\phi}(t_{1})=\delta\dot{\phi}(t_{2})=0\,.
\end{equation}
These conditions imply that the gauge function $\epsilon(t)$ should vanish at the boundaries along with its time derivative $\dot\epsilon(t)$. Hence, one cannot make a residual gauge transformation to eliminate $c_0$. Therefore, the number of initial data needed to specify the motion is four and the number of DoF is two. We will confirm this conclusion in the next section. 
But before going there, it is important to stress that the presence of time derivatives in the transformation \eqref{eq: ex1: transformation} is only important if accompanied with the singularity of the Jacobian. Indeed, one can easily check that transformation 
\begin{equation}
    q_1(Q,\dot{\phi})=Q+\dot{\phi}\ ,\label{eq: regular transformation_deriv}
\end{equation}
neither changes the number of DoF nor dynamics. 
On the other hand, singular transformations without time derivatives are also trivial and do not change dynamics. For instance, this is the case for
\begin{equation}
    q_2(Q,\phi)=Q^3+\phi\ .\label{eq: regular transformation_No_deriv}
\end{equation}
The transformed action \eqref{eq: ex1: first transformed action} clearly describes a higher derivative theory in the field $\phi$. Therefore our first step is to reduce the order of derivatives by introducing a constraint so that the Lagrangian becomes 
\begin{equation}
    L=\frac{1}{2}\left(3Q^{2}\dot{Q}+\dot{\theta}\right )^{2}+\lambda(\dot{\phi}-\theta)\ .
    \label{eq: action_lower_der}
\end{equation}
Now the gauge transformations \eqref{eq: ex1: gauge freedom} become
\begin{equation}
\phi\rightarrow\phi+\epsilon\,,\quad\:Q^{3}\rightarrow Q^{3}-\dot{\epsilon}\,,\quad\:\theta\rightarrow\theta+\dot{\epsilon}\,.
\end{equation}
We can introduce the gauge invariant variable
\begin{equation}
\bar{q}=Q^{3}+\theta\,,
\end{equation}
using which the action \eqref{eq: action_lower_der} takes the form 
\begin{equation}
L=\frac{1}{2}\dot{\bar{q}}^{2}+\lambda\left(\dot{\phi}+Q^{3}-\bar{q}\right)\,.\label{eq:L_with_Q}
\end{equation}
Now the equation of motion for the auxiliary variable $Q$ 
\begin{equation}
\lambda\,Q=0\,,
\end{equation}
introduces two branches. Following \cite{Pons:2009ch} we can substitute $Q=0$ back into the action to find the action for the singular branch, so that the Lagrangian \eqref{eq:L_with_Q} takes the form:
\begin{equation}
L=\frac{1}{2}\dot{\bar{q}}^{2}+\lambda\left(\dot{\phi}-\bar{q}\right)\,.\label{eq:L_without_Q}
\end{equation} 
Note that the singular branch includes the solutions $q(t)$ of the regular branch.
This structure is similar to the covariant unimodular gravity \cite{Henneaux:1989zc} and unicurvature
gravity \cite{Jirousek:2020vhy}, as $\dot{\phi}$ enters only linearly and $\lambda$
is the canonical momentum for $\phi$. This theory models acceleration
as a Lagrange multiplier or constant of integration. 
An important
difference between actions \eqref{eq:L_with_Q} and \eqref{eq:L_without_Q}
is that in the first action $Q=0$ picks up the singular branch with the novel solutions while in the second action it is $\lambda=0$
which corresponds to the regular branch with original solutions of the seed theory. Crucially, to pick up $Q=0$ in theory given by \eqref{eq:L_with_Q} corresponds to zero measure. Thus, in most cases the dynamics describe a free particle and the accelerated particle appears as a very singular improbable configuration. On the other hand, to pick up $\lambda=0$ in theory given by \eqref{eq:L_without_Q} corresponds to measure zero. So that the original dynamics of a free particle appear to be a singular improbable case.

Finally, we would like to highlight the structure of the action \eqref{eq:L_with_Q}. This expression clearly corresponds to the original action \eqref{eq: ex1: action}, with the transformation \eqref{eq: ex1: transformation} enforced through a Lagrange multiplier. It turns out that we can implement any transformation in this manner as long as the original variables are given explicitly as functions of the new variables i. e.
\begin{equation}
    q^{i}=q^{i}(Q^{a})\ .\label{eq: ex1: generalized transformation}
\end{equation}
The lowercase $q^{i}$ denotes the old variables here, while capital $Q^{a}$ corresponds to the new ones. Note that the number of novel variables does not need to match the number of old ones, hence, the different index $i$ and $a$. If we consider applying this transformation to a seed Lagrangian $\mathcal{L}(q^{i},\dot{q}^{j})$, we can equivalently write such theory using a constraint as
\begin{equation}
    S[q^{i},Q^{a},\lambda_{j}]=\int dt\,\Big [\mathcal{L}(q^{i},\dot{q}^{j})+\lambda_{i}\left(q^{i}-q^{i}(Q^{a})\right)\Big ]\ .
\end{equation}
To see that this theory indeed corresponds to the transformed theory, we can use the constraint to plug in for $q^{i}$ in the original part of the Lagrangian $\mathcal{L}(q^{i},\dot{q}^{j})$. After that $q^{i}$ appears only linearly in the constraints and thus it becomes a Lagrange multiplier enforcing $\lambda_{i}=0$. This allows us to integrate it out of the action so that we are left with just the original Lagrangian subjected to the transformation \eqref{eq: ex1: generalized transformation}.

\subsection{Hamiltonian analysis}
\label{sec:Ham toy}
Let as count the DoF in the transformed theory \eqref{eq: ex1: first transformed action} using full-fledged canonical analysis applied to the system with action \eqref{eq: action_lower_der}. There, the canonical momenta are 
\begin{align}
    p_{\theta}=&\dot{q}\ ,\label{eq: Ham1: non-trivial momentum}\\
    p_{Q}=&3Q^{2}\,\dot{q}\ ,\\
    p_{\phi}=&\lambda\ ,\label{eq: Ham1: primary constraint phi}\\
    p_{\lambda}=&0\ ,\label{eq: Ham1: primary constraint lambda}
\end{align}
where $\dot{q}=3Q^{2}\dot{Q}+\dot{\theta}$. The last two expressions clearly represent primary constraints. However, substituting the first equality into the second to eliminate $\dot{q}$ reveals an additional primary constraint in the theory
\begin{equation}
    p_{Q}=3Q^{2}\,p_{\theta}\ .\label{eq: Ham1: primary constraint Q}
\end{equation}
Hence, we have only a single unconstrained momentum $p_{\theta}$. Note that we are therefore unable to solve for both $\dot{\theta}$ and $\dot{Q}$ separately. Nevertheless the relation \eqref{eq: Ham1: non-trivial momentum} is enough to eliminate the dependence on all dotted variables in the Hamiltonian. Indeed, after a bit of algebra one can find the Dirac Hamiltonian
\begin{equation}
    H_{D}=\frac{1}{2}p_{\theta}^{2}+p_{\phi}\theta+\alpha\left (p_{Q}-3Q^{2}p_{\theta}\right )+\beta\, p_{\lambda}+\gamma\,(p_{\phi}-\lambda)\ ,
\end{equation}
where $\alpha$, $\beta$ and $\gamma$ are Lagrange multipliers enforcing primary constraints. Note that we have used the primary constraints \eqref{eq: Ham1: primary constraint phi} to \eqref{eq: Ham1: primary constraint Q} to get this form. The consistency condition for the primary constraints \eqref{eq: Ham1: primary constraint phi} and \eqref{eq: Ham1: primary constraint lambda} allows us to determine the multipliers $\beta$ and $\gamma$:
\begin{align}
    \{p_{\phi}-\lambda,H_{D}\}=&\,-\beta=0\ ,\label{eq: Ham1: consistency phi}\\
    \{p_{\lambda},H_{D}\}=&\,\gamma=0\ .\label{eq: Ham1: consistency lambda}
\end{align}
Hence, the Hamiltonian simplifies to
\begin{equation}
    H_{D}=\frac{1}{2}p_{\theta}^{2}+p_{\phi}\theta+\alpha\left (p_{Q}-3Q^{2}p_{\theta}\right )\ .
\end{equation}
Finally, the consistency condition induced by the last primary constraint \eqref{eq: Ham1: primary constraint Q} gives
\begin{equation}
   \{p_{Q}-3Q^{2}\,p_{\theta},H_{D}\}=3 Q^{2}p_{\phi}=0\ .\label{eq: Ham1: branching consistency condition}
\end{equation}
The above condition is a product of two factors, which reflects the branching nature of the theory. This product structure of the constraint implies that the constraint surface is \emph{not a differentiable manifold}. Rather it consists of two manifolds, corresponding to the two branches of solutions, with a non-trivial intersection\footnote{This situation can be easily illustrated in $\mathbb{R}^{2}$. Suppose we are solving the equation $xy=0$. The solution space here is clearly the union of the $x$ axis and the $y$ axis. While both axes are manifolds themselves, their union is not a manifold due to the non-trivial intersection at $x=y=0$.}. This clearly goes beyond the standard assumptions \cite{Henneaux:1992ig} for analysis of constrained systems. We proceed with the analysis by studying each of these manifolds separately. Hence, we either impose the secondary constraint $Q=0$, which clearly corresponds to the singular point of the Jacobian \eqref{eq: ex1: Jacobian}, or we take
\begin{equation}
  p_{\phi}=0 \ ,\label{eq: zero p_phi}
\end{equation}
which, as we shall demonstrate, corresponds to the regular branch.\par

We first focus on the regular branch by introducing the secondary constraint \eqref{eq: zero p_phi}. This constraint commutes with all the primary constraints \eqref{eq: Ham1: primary constraint phi} to \eqref{eq: Ham1: primary constraint Q} and with the Hamiltonian itself. Hence, we do not generate tertiary constraints. Furthermore, this means that \eqref{eq: zero p_phi} is a first class constraint. Analysing the commutation relations of the primary constraint we find that \eqref{eq: Ham1: primary constraint Q} is also first class, while \eqref{eq: Ham1: primary constraint phi} and \eqref{eq: Ham1: primary constraint lambda} are second class. Since the phase space is 8 dimensional we get the following degree of freedom count
\begin{equation}
    \mathrm{\# D.o.F.}=\frac{8-2\times 2-2\times 1}{2}=1\ .
\end{equation}
To demonstrate that this branch indeed corresponds to the phase space of the original theory we can directly confirm that $\ddot{q}=0$:
\begin{align}
    \dot{q}=\,\{q,H_{D}\}=&\,p_{\theta}\ ,\\
    \ddot{q}=\,\{\dot{q},H_{D}\}=&\,p_{\phi}=0\ ,\label{eq: Ham1: q eom}
\end{align}
where in the last line we use the secondary constraint $p_{\phi}=0$. Finally, since \eqref{eq: Ham1: primary constraint Q} is a first class constraint, the associated multiplier $\alpha$ remains in the Hamiltonian. This implies that there is a gauge symmetry in our description, which is generated by the constraint \eqref{eq: Ham1: primary constraint Q}. However, we can confirm that both $q=Q^{3}+\theta$ and the associated $p_{q}=\dot{q}=p_{\theta}$ are invariant under this symmetry since they commute with the constraint \eqref{eq: Ham1: primary constraint Q}
\begin{align}
    \delta_{\epsilon'}q\equiv&\epsilon'\{q,p_{Q}-3Q^{2}p_{\theta}\}=0\ ,\\
    \delta_{\epsilon'}p_{q}\equiv&\epsilon'\{p_{q},p_{Q}-3Q^{2}p_{\theta}\}=0\ .\label{eq: Ham1: gauge symmetry}
\end{align}

Let us now return and analyze the singular branch. Here the secondary constraint is $Q=0$, while $p_{\phi}$ is left undetermined. The crucial difference in comparison to the regular branch is that $Q=0$ is not a first class constraint. This can be seen from its non-vanishing commutator with \eqref{eq: Ham1: primary constraint Q}. In turn \eqref{eq: Ham1: primary constraint Q} becomes a second class constraint as well. Consequently, the consistency condition for $Q=0$ allows us to determine the last multiplier $\alpha$:
\begin{equation}
    \{Q,H_{D}\}=\alpha=0\ .
\end{equation}
Therefore, we get the Hamiltonian
\begin{equation}
    H_{D}=\frac{1}{2}p_{\theta}^{2}+p_{\phi}\theta\ .
\end{equation}
Here, it should be noticed that contrary to the regular branch the Hamiltonian depends linearly on the momentum $p_\phi$, that is different from the case of the regular branch with $p_\phi=0$, which is the manifestation of the Ostrogradsky ghost\footnote{The Ostrogradsky ghost is usually considered to imply instability, but there is a recent exception \cite{Deffayet:2021nnt} from this common lore.}, see \cite{Ostrogradsky:1850fid,Woodard:2006nt}.
If one would like to avoid the appearance of such a ghost degree of freedom, one can consider the model with $L = q^2/2$, for example, in which there is no dynamics in the regular branch, but a new healthy degree of freedom appears in the singular branch. Finally, the fact that $Q=0$ and \eqref{eq: Ham1: primary constraint Q} are second class is reflected in the number of degrees of freedom in the theory, which is now
\begin{equation}
    \mathrm{\# D.o.F.}=\frac{8-4\times 1}{2}=2\ ,
\end{equation}
and clearly has doubled in comparison with the regular branch. This can be easily traced to the consistency condition \eqref{eq: Ham1: branching consistency condition}, where the Jacobian \eqref{eq: ex1: Jacobian} enters as an overall factor. Hence, \emph{any} singular point of the Jacobian can be taken as a novel branch of the theory. Crucially, this allows us to evade the constraint $p_{\phi}=0$ opening the possibility of non-vanishing second derivative of $q$. Instead we get
\begin{equation}
    \dddot{q}=\{p_{\phi},H_{D}\}=0\ .
\end{equation}
Lastly, let us note that the gauge symmetry \eqref{eq: Ham1: gauge symmetry} no longer exists in the singular branch, since its action does not preserve the secondary constraint $Q=0$. This is in agreement with our observation that the symmetry \eqref{eq: ex1: gauge freedom} cannot be applied to the singular branch.

\subsection{Non-trivial kinetic term}\label{Non-trivial kin.term}
The novel degree of freedom we have found in the last section is clearly tied to the newly introduced variable $\phi$, which has not appeared in the action prior to transformation \eqref{eq: ex1: transformation}. It is therefore natural to ask, whether the number of DoF will still increase even when a degree of freedom associated to $\phi$ has already existed in the theory before we introduce it through \eqref{eq: ex1: transformation}. In fact, many interesting mimetic theories and equivalents e.g. \cite{Chamseddine:2014vna,Mirzagholi:2014ifa,Capela:2014xta,Lim:2010yk,Arroja:2015wpa,Ramazanov:2015pha} do have the seed action explicitly dependent on field $\phi$ facilitating the transformation. We explore this option by modifying the toy model \eqref{eq: ex1: action} by including a kinetic term for the variable $\phi$.
\begin{equation}
    S[q,\phi]=\frac{1}{2}\int_{t_{1}}^{t_{2}} dt\,\left (\dot{q}^{2}+\dot{\phi}^{2}\right )\ .\label{eq: ex2: action}
\end{equation}
Let us now consider the same transformation \eqref{eq: ex1: transformation} applied to the above action. Since $\phi$ already exists in the theory we have to extend this transformation to $\phi$. We chose the following trivial extension
\begin{align*}
    q=&\,Q^{3}+\dot{\phi}\ ,\\
    \phi =&\,\phi\ .\numberthis{}\label{eq: ex2: transformation}
\end{align*}
After this transformation the action becomes
\begin{equation}
    S\left[Q,\phi\right]=\frac{1}{2}\int_{t_{1}}^{t_{2}}dt\,\left [ \left(\ddot{\phi}+3Q^{2}\dot{Q}\right)^{2}+\dot{\phi}^{2}\right ]\,,\label{eq: ex1: transformed action}
\end{equation}
and the equations of motion associated to $Q$ and $\phi$ respectively are
\begin{align}
    \Ddot{q}\,Q^{2}=&\,0\ ,\label{eq: ex2: Q eom}\\
    \frac{d}{dt}\left [\Ddot{q}-\dot{\phi}\right]=&\,0\ .\label{eq: ex2: phi eom}
\end{align}
We can immediately see that the structure of the first equation remains unchanged. We again obtain two branches in the solution space - the regular branch ($\ddot{q}=0$) and the singular one ($Q=0$). In the regular branch the $\phi$-equation of motion is, however, no longer empty and implies linear evolution for the variable $\phi$
\begin{equation}
    \ddot{\phi}=0\ .
\end{equation}
This is clearly the $\phi$ equation of motion of the original theory \eqref{eq: ex2: action}. Hence, we see that the original dynamics of action \eqref{eq: ex1: action} is recovered in the regular branch. Unlike in \cref{sec:toy}, we now have two degrees of freedom instead of one due to $\phi$ being dynamical to begin with. This can be further tied to the fact that the gauge symmetry \eqref{eq: ex1: gauge freedom}, which can be held responsible for the missing degree of freedom, is manifestly broken due to the addition of the kinetic term for $\phi$. The singular branch is given by the vanishing of the Jacobian \eqref{eq: ex1: Jacobian}, which simplifies relation \eqref{eq: ex2: transformation} to $q=\dot{\phi}$. Hence, $\phi$-equation of motion \eqref{eq: ex2: phi eom} becomes
\begin{equation}
    \frac{d^{2}}{dt^{2}}\left [\Ddot{\phi}-\phi\right]=\,0\ ,
\end{equation}
giving us the solution for $\phi$
\begin{equation}
    \phi=c_{0}+c_{1}t+c_{2}\,\mathrm{cosh}(t)+c_{3}\,\mathrm{sinh}(t)\ .
\end{equation}
Here $c_{i}$ are 4 constants of integration, suggesting that there are again 2 degrees of freedom in the singular branch. We see that there was no increase in the number of degrees of freedom between the two branches this time. Nevertheless, the solutions encountered in the singular branch are significantly altered in comparison to the solutions of the original theory. This is true for the variable $q$ as well as for $\phi$
\begin{equation}
    q=\dot{\phi}=c_{1}+c_{2}\,\mathrm{sinh}(t)+c_{3}\,\mathrm{cosh}(t)\ .
\end{equation}
\par

We proceed to the Hamiltonian analysis in this modified example to confirm our naive counting of the degrees of freedom. We again introduce the variable $\theta$ in order to reduce the order of derivatives
\begin{equation}
    L\rightarrow\frac{1}{2}\dot{q}(Q,\theta)^{2}+\frac{1}{2}\theta^{2}+\lambda(\dot{\phi}-\theta)\ ,
\end{equation}
and convert the kinetic term for $\phi$ to a potential term for $\theta$. Since the modification now appears only as a potential term much of the analysis remains unaffected by its addition. In particular, we obtain only one non-constrained canonical momentum
\begin{equation}
    p_{\theta}=\dot{q}\ ,
\end{equation}
along with three primary constraints
\begin{align}
    p_{\phi}=&\lambda\ ,\label{eq: Ham2: primary constraint phi}\\
    p_{\lambda}=&0\ ,\label{eq: Ham3: primary constraint lambda}\\
    p_{Q}=&3Q^{2}\,p_{\theta}\ .\label{eq: Ham2: primary constraint Q}
\end{align}
The Hamiltonian is therefore only modified by the potential term $\theta^{2}/2$
\begin{equation}
    H_{D}=\frac{1}{2}p_{\theta}^{2}+p_{\phi}\theta-\frac{1}{2}\theta^{2}+\alpha\left (p_{Q}-3Q^{2}p_{\theta}\right )+\beta\, p_{\lambda}+\gamma\,(p_{\phi}-\lambda)\ .
\end{equation}
The consistency conditions for the constraints \eqref{eq: Ham2: primary constraint phi} and \eqref{eq: Ham3: primary constraint lambda} also remain unchanged and therefore we get $\beta=0$ and $\gamma=0$ as in \eqref{eq: Ham1: consistency phi} and \eqref{eq: Ham1: consistency lambda}. So the Hamiltonian becomes
\begin{equation}
    H_{D}=\frac{1}{2}p_{\theta}^{2}+p_{\phi}\theta-\frac{1}{2}\theta^{2}+\alpha\left (p_{Q}-3Q^{2}p_{\theta}\right )\ .
\end{equation}
Finally, the consistency condition for primary constraint \eqref{eq: Ham2: primary constraint Q} is given by 
\begin{equation}
   \{p_{Q}-p_{\theta}3Q^{2},H_{D}\}=3 Q^{2}(p_{\phi}-\theta)=0\ ,
\end{equation}
and is different from \eqref{eq: Ham1: branching consistency condition}. Like in \cref{sec:Ham toy} this consistency requirement introduces a secondary constraint to the theory, which is given as a product of two factors. These correspond to the two branches - regular and singular - as we have identified them in our analysis of equation of motion \eqref{eq: ex2: Q eom}. The secondary constraint characterizing the singular branch remains unchanged as it is given by $Q=0$. However, the constraint corresponding to the regular branch obtains an additional term $\theta$ in comparison to \eqref{eq: zero p_phi}:
\begin{equation}
     p_{\phi}-\theta=0\ .\label{eq: Ham2: secondary constraint regular}
\end{equation}
This novel term in the constraint changes the commutation relation between \eqref{eq: Ham2: secondary constraint regular} and \eqref{eq: Ham2: primary constraint Q}, which does not vanish now. This means that the secondary constraint in the regular branch is second class and we can use it to determine the multiplier $\alpha$. In particular we get the following consistency condition
\begin{equation}
    \{p_{\phi}-\theta,H_{D}\}=-p_{\theta}+\alpha\, 3Q^{2}\ .
\end{equation}
Hence, the Hamiltonian becomes
\begin{equation}
    H_{D}=\frac{1}{2}p_{\theta}^{2}+p_{\phi}\theta-\frac{1}{2}\theta^{2}+p_{\theta}\left (\frac{p_{Q}}{3Q^{2}}-p_{\theta}\right )\ .
\end{equation}
Moreover, unlike in \cref{sec:Ham toy} the secondary constraint is second class even in the regular branch. This results in a higher number of degrees of freedom. In particular, we get
\begin{equation}
    \mathrm{\# D.o.F.}=\frac{8-4\times 1}{2}=2\ .
\end{equation}
In the singular branch the constraint structure is exactly the same as in the singular branch from \cref{sec:Ham toy}. In particular the secondary constraint $Q=0$ is second class as it does not commute with \eqref{eq: Ham2: primary constraint Q}. Hence, we obtain the following consistency condition
\begin{equation}
    \{Q,H_{D}\}=\alpha=0\ .
\end{equation}
The final Hamiltonian is thus only changed by the $\theta$ potential
\begin{equation}
    H_{D}=\frac{1}{2}p_{\theta}^{2}+p_{\phi}\theta-\frac{1}{2}\theta^{2}\ .
\end{equation}
Consequently, we get 2 degrees of freedom as expected.

\section{Even More Extra Degrees of Freedom for  Relativistic Free Particle}
\label{sec:relativistic}

So far we have demonstrated that the dynamics originating in the singular point of the Jacobian \eqref{eq: ex1: Jacobian} can introduce a single novel degree of freedom. This appearance could be traced not only to the singular point but also to the appearance of $\dot{\phi}$ in the transformation \eqref{eq: ex1: transformation}. In this section, we aim to demonstrate that the amount of degrees of freedom introduced to the theory due to an invertible transformation featuring $\dot{\phi}$ may be larger then one. We show this on the model for free relativistic particle of unit mass in n-dimensional Minkowski background\footnote{We use the signature convention $\left(+,-,-,-\right)$.} described by a polynomial time-reparametrisation invariant action 
\begin{equation}
    S[x^{\mu},e]=-\frac{1}{2}\int\,d\tau\ \Big (\frac{\dot{x}^{2}}{e}+e\Big )\ .\label{eq:einbein theory}
\end{equation}
Here $x^{\mu}$ are the coordinates of the particle trajectory in spacetime, $\tau$ is the time coordinate on the world-line and $e$ is the einbein. We are using a shorthand $\dot{x}^{2}=\eta_{\mu\nu}\dot x^{\mu}\dot x^{\nu}$. The equations of motion for the coordinate variables are
\begin{equation}
    \frac{d}{d\tau}\frac{\dot{x}^{\mu}}{e}=0\ .\label{eq:x eom}
\end{equation}
The equation of motion associated with the einbein $e$ is
\begin{equation}
    \frac{\dot{x}^{2}}{e^{2}}=1\ .
\end{equation}
which we can solve for the einbein directly to obtain
\begin{equation}
    e=\sqrt{\dot{x}^{2}}\ .\label{eq:g eom}
\end{equation}
Plugging this into \eqref{eq:x eom} we obtain
\begin{equation}
    \frac{d}{d\tau}\frac{\dot{x}^{\mu}}{\sqrt{\dot{x}^{2}}}=0\ .
\end{equation}
From this we can see that the theory in is particularly simple when $n=1$ as there are no dynamics whatsoever in such case.

The field $e$ is clearly just an auxiliary variable that can be solved from its own equation of motion algebraically. Thus we can integrate it out by plugging the solution back into the action \eqref{eq:einbein theory}. This returns the standard non-polynomial action for the free particle
\begin{equation}
    S[x^{\mu}]=-\int\,d\tau\sqrt{\dot{x}^{2}}\,\ .
\end{equation}
Before we consider transformations of this theory let us work out the Hamiltonian analysis of this system. The canonical momenta corresponding to $x^{\mu}$ and $e$ respectively are
\begin{equation}
    p_{\mu}=\frac{\partial \mathcal{L}}{\partial \dot{x}^{\mu}}=-\frac{\dot{x}_{\mu}}{e}\ ,\qquad\mathrm{and}\qquad \pi=\frac{\partial \mathcal{L}}{\partial \dot{e}}=0\ .
\end{equation}
We see that we have one primary constraint $\pi=0$. The Dirac Hamiltonian is therefore
\begin{equation}
    H_{D}=\frac{e}{2}(1-p^{2})+\lambda \pi\ .\label{eq:einbein dirac Hamiltonian}
\end{equation}
Here $\lambda$ is an undetermined multiplier and $p^{2}=\eta^{\mu\nu}p_{\mu}p_{\nu}$. The consistency condition associated with the primary constraint reads
\begin{equation}
    \{\pi,H_{D}\}=\frac{1}{2}(p^{2}-1)=0\ .\label{eq: ein: secondary constraint}
\end{equation}
Hence, we obtain a secondary constraint $p^{2}=1$, which commutes with the Hamiltonian
\begin{equation}
    \{p^{2}-1,H_{D}\}=0\ .
\end{equation}
The two constraints commute with each other
\begin{equation}
    \{p^{2}-1,\pi\}=0\ ,
\end{equation}
hence, we have two first class constraints. Since we have started with $n+1$ independent variables the dimensionality of the phase space is $2(n+1)$. The first class constraints are subtracted twice so we have the final number of degrees of freedom being $n-1$. In the special case of $n=1$ this returns $0$ which agrees with our intuition from the equations of motion.

\subsection{Einbein Transformation}
\label{sec: ein transformation}
Let us construct a transformation of the einbein $e$ in hopes that we can increase the D.o.F. count of the above theory. In order to do so, we consider a novel einbein $f$ and an additional scalar $\phi$ to facilitate the following relationship
\begin{equation}
    e=(f-\dot{\phi})^{3}+\dot{\phi}^{3}=f\left(3\dot{\phi}^{2}-3f\dot{\phi}+f^{2}\right)\ ,\label{eq:einbein transfromation}
\end{equation}
where expression in the brackets on the right hand side emulates the gauge transformations of the einbein due to the time-reparametrisation on the world-line. 
Note that the transformation also maps the field $\phi$ to itself
\begin{equation}
    \phi=\phi\ .
\end{equation}
The transformation of the field $\phi$ is trivial and therefore the invertibility of the entire transformation depends only on the map between the two einbeins \eqref{eq:einbein transfromation}, which is in our case indeed invertible. This can be seen directly by solving for $f$ as
\begin{equation}
    f=\sqrt[3]{e-\dot{\phi}^{3}}+\dot{\phi}\ .
\end{equation}
Despite this, the transformation is not regular everywhere. In fact it has a vanishing Jacobian at $f=\dot{\phi}$. For the Jacobian we get
\begin{equation}
    J=\frac{\partial e}{\partial f}=3(f-\dot{\phi})^{2}\ .\label{eq:einbein jacobian}
\end{equation}
Correspondingly, the inverse transformation is non-differentiable at $e=\dot{\phi}^{3}$. We now apply the transformation \eqref{eq:einbein transfromation} to the theory \eqref{eq:einbein theory}. Let us first take a look at the equations of motion of the transformed theory. The equations for the fields $x^{\mu}$ remain unchanged and read
\begin{equation}
    \frac{d}{d\tau}\frac{\dot{x}_{\mu}}{e}=0\ .\label{eq: ein: x eom}
\end{equation}
On the other hand, the equation of motion for $f$ consists of two multiplicative factors
\begin{equation}
    \left (1-\frac{\dot{x}^{2}}{e^{2}}\right )(f-\dot{\phi})^{2}=0\ .\label{eq:h eom}
\end{equation}
The first term clearly corresponds to the original equation of motion for $e$ \eqref{eq:g eom}, while the second factor corresponds to the Jacobian of the transformation \eqref{eq:einbein transfromation}. Hence, we have two branches of solutions - the regular branch containing the original solutions \eqref{eq:g eom} or the singular branch where the Jacobian \eqref{eq:einbein jacobian} vanishes
\begin{equation}
    f=\dot{\phi}\ .\label{eq: ein: singular branch}
\end{equation}
In the regular branch the equation of motion for $\phi$ is satisfied automatically
\begin{equation}
    \frac{d}{d\tau}\Bigg [\left (1-\frac{\dot{x}^{2}}{e^{2}}\right )\,f\,(2\dot{\phi}-f)\Bigg]=0\ .\label{eq:phi eom}
\end{equation}
On the other hand in the singular branch it can be used to solve the evolution of the factor $1-\dot{x}^{2}/e^{2}$, which remains undetermined from equation \eqref{eq:h eom}. We can plug \eqref{eq: ein: singular branch} to simplify this equation to obtain\footnote{Note that for $f=0$ this equation is also satisfied trivially even in the singular branch. This prevents us from solving for the coordinate variables $x^{\mu}$, which consequently become pure gauge. We will disregard this possibility as it clearly corresponds to the case $e=0$ for which the original action is clearly ill-defined.}
\begin{equation}
    \frac{d}{d\tau}\Bigg [\left (1-\frac{\dot{x}^{2}}{e^{2}}\right )\,f^{2}\Bigg]=0\ .\label{eq:phi eom singular}
\end{equation}
This is clearly analogical to what happened with $\ddot{q}$ in \cref{sec:toy}. The crucial difference between the two examples is that $1=\dot{x}^{2}/{e^{2}}$ is a constraint \eqref{eq: ein: secondary constraint} in the original theory \eqref{eq:einbein theory}, whereas $\ddot{q}=0$ is just an equation of motion. Hence, uplifting this constraint not only introduces extra freedom in the theory but it further affects the constraint structure. This, as we will see, will give way for additional degrees of freedom to appear.
We first show that this indeed occurs by solving the equations of motion in the singular branch for the simplest case $n=1$. The equation of motion \eqref{eq: ein: x eom} can be easily integrated as
\begin{equation}
    \frac{\dot{x}}{e}=c_{1}\ ,
\end{equation}
where $c_{1}$ is a constant of integration. Plugging this into \eqref{eq:phi eom singular} we can likewise find
\begin{equation}
    f=c_{2}=const.
\end{equation}
Plugging this into \eqref{eq: ein: singular branch} we obtain
\begin{equation}
    \phi=c_{2}\tau+c_{3}\ ,
\end{equation}
where $c_{3}$ is a constant. For the original einbein we get
\begin{equation}
    e=c_{2}^{3}\ ,
\end{equation}
which yields the solution for $x$
\begin{equation}
    x=\frac{c_{1}}{c_{2}^{3}}\tau+c_{4}\ .
\end{equation}
We see that we obtained $4$ constants of integration $c_{i}$, while the original theory had none. Hence, we expect there to be 2 more degrees of freedom in the transformed theory for the general $n$. Note that for $f=c_{2}=0$, $x$ is not well defined and thus leads to a break down of the theory. This again corresponds to $e=0$, which is clearly not a physical configuration.

\subsection{Hamiltonian analysis}
Let us confirm the above estimate by performing Hamiltonian analysis of the full theory \eqref{eq:einbein theory} in $n$ dimensions transformed using \eqref{eq:einbein transfromation}. The canonical momenta are
\begin{align}
    p_{\mu}&=\frac{\partial \mathcal{L}}{\partial \dot{x}^{\mu}}=-\frac{\dot{x}_{\mu}}{e}\ ,\label{eq: ein: x momentum}\\
    \pi&=\frac{\partial \mathcal{L}}{\partial \dot{f}}=0\ ,\label{eq: ein: primary constraint}\\
    p_{\phi}&=\frac{\partial \mathcal{L}}{\partial \dot{\phi}}=\frac{3f}{2}(p^{2}-1)(2\dot{\phi}-f)\ ,\label{eq:phi momentum}
\end{align}
where we have solved for $\dot{x}^{\mu}$ using $p_{\mu}$ in the last line. At first glance this theory has a single primary constraint $\pi=0$; however, it is crucial to note that the last equation is only solvable for $\dot{\phi}$ when $p^{2}\neq 1$. This is again a non-standard situation, which goes beyond the standard assumptions for the Hamiltonian analysis \cite{Henneaux:1992ig}. Namely, the rank of the Jacobian of the map between the momenta and velocities \eqref{eq: ein: x momentum} to \eqref{eq:phi momentum} is not constant throughout the configuration space. It is lower when $p^{2}=1$.\footnote{Note that the same situation occurs for the unphysical configurations $f=0$.} Clearly, we cannot continue with the analysis using the standard procedure as it is unclear what is our primary constraint structure. For this reason we analyze both cases separately. Clearly the branching in this theory occurs much earlier in the analysis than in the examples from \cref{sec:toy}. We first focus on the case $p^{2}\neq1$. We expect that this case will correspond to the singular branch of the theory as the original constraint \eqref{eq: ein: secondary constraint} is explicitly violated here. The Dirac Hamiltonian in this case has the following form
\begin{equation}
    H_{D}=\frac{1}{2}e(1-p^{2})+p_{\phi}\dot{\phi}+\lambda \pi\ .
\end{equation}
Here $\lambda$ is an undetermined multiplier associated with the primary constraint $\pi=0$. Note that we have deliberately opted to keep $\dot{\phi}$ and $e$ explicit in order to increase the readability of the equation. However, $\dot{\phi}$ is meant to be solved from \eqref{eq:phi momentum} as
\begin{equation}
    \dot{\phi}=\frac{1}{2}\left [\frac{2p_{\phi}}{p^{2}-1}\frac{1}{3f}+f\right]\ ,
\end{equation}
while $e$ is given by \eqref{eq:einbein transfromation}.
The consistency condition associated with the primary constraint \eqref{eq: ein: primary constraint} gives
\begin{equation}
    \{\pi,H_{D}\}=\frac{3}{2}(f-\dot{\phi})^{2}(p^{2}-1)=0\ .
\end{equation}
The structure of this secondary constraint is the same as \eqref{eq: Ham1: branching consistency condition} as there are two factors multiplied with each other. The difference here is that we have explicitly assumed that $p^{2}\neq 1$ and therefore we get just a single constraint
\begin{equation}
    \dot{\phi}=f\ .\label{eq: ein: secondary constraint singular}
\end{equation}
This condition clearly corresponds to the vanishing of the Jacobian \eqref{eq:einbein jacobian}. Hence, the singular point of the Jacobian again appears as a secondary constraint of the theory. Furthermore, this constraint does not commute with \eqref{eq: ein: primary constraint}. Instead we get
\begin{equation}
    \{\pi,\dot{\phi}-f\}=\frac{\dot{\phi}}{f}=1\ ,\label{eq:constraint commutator}
\end{equation}
which implies that both constraints are second class. Consequently, the consistency condition associated with \eqref{eq: ein: secondary constraint singular} allows us to determine $\lambda$ as
\begin{equation}
    \{f-\dot{\phi},H_{D}\}=\lambda=0\,.
\end{equation}
Hence, the Hamiltonian becomes
\begin{equation}
    H_{D}=\frac{1}{2}e(1-p^{2})+p_{\phi}\dot{\phi}\ .
\end{equation}
There are no additional constraints in the theory and we can proceed to calculate the number of degrees of freedom. We began with $n$ variables corresponding to the coordinates $x^{\mu}$ along with the einbein $f$ and the variable $\phi$. Hence, we have $n+2$ variables producing a $2n+4$ dimensional phase space. We obtained $2$ second class constraints giving us a total of $n+1$ degrees of freedom. This is clearly $2$ more than the original theory, which confirms our guess from solving the equations of motion.



Let us now return to the second case, where we restrict ourselves to $p^{2}=1$. This is clearly a constraint of the original theory \eqref{eq:einbein theory} and therefore we expect that the associated dynamics will correspond to the regular branch of the theory. The above restriction immediately implies that $p_{\phi}=0$, due to the relation \eqref{eq:phi momentum}. We proceed with the analysis by interpreting both of these expressions as primary constraints of the theory. Hence, we obtain the Dirac Hamiltonian
\begin{equation}
    H_{D}=\lambda_{1}(1-p^{2})+\lambda_{2}\pi+\lambda_{3}p_{\phi}\ ,\label{eq: ein: Hamiltonian regular}
\end{equation}
where $\lambda_{1,2,3}$ are undetermined multipliers. All the primary constraints commute with each other and, by extension, with the above Hamiltonian. Therefore, we do not obtain any further secondary constraints and we find that all the constraints are first class. Hence, the degree of freedom count is $n-1$ as in the original theory. Despite the matching DoF count, the resulting Hamiltonian \eqref{eq: ein: Hamiltonian regular} appears to differ from \eqref{eq:einbein dirac Hamiltonian}. However, we can easily see that the dynamics are indeed equivalent. In the original theory the eibein $e$ has the role of a metric on the world-line and it can be chosen arbitrarily\footnote{Up to the unphysical choice $e=0$.} due to the associated time-reparametrisation invariance on the world-line. This freedom is captured here directly by the free parameter $\lambda_{1}$, which plays the same role in equation \eqref{eq: ein: Hamiltonian regular} as einbein $e$ in \eqref{eq:einbein dirac Hamiltonian}. Alternatively we can recover \eqref{eq:einbein dirac Hamiltonian} by choosing $\lambda_{1}=e/2$. Finally, there is an extra term $\lambda_{3}\,p_{\phi}$, which does not affect the dynamics of other variables.


\section{Example of Invertible Disformal Transformation resulting in Mimetic DM}
\label{sec:mimetic}
As we have seen the appearance of the novel branch of solutions in the transformed theory is intimately tied to the existence of singular points of the Jacobian of the transformation. In fact, we have utilized this exact mechanic in our recent paper \cite{Jirousek:2022rym}, where we have demonstrated that singular points of disformal transformations give rise to novel branches of solutions with mimetic dark matter \cite{Chamseddine:2013kea}. The regular disformal transformations are often employed in recent studies of modified gravity, see e.g. \cite{Bettoni:2013diz,Domenech:2015tca,Crisostomi:2016czh,BenAchour:2016cay,Crisostomi:2016tcp,Takahashi:2017pje,Deffayet:2020ypa,Dusoye:2020wom}. In this section we aim to show that the results of \cite{Jirousek:2022rym} are applicable even to disformal transformations which are locally or even globally invertible. We do this by directly constructing an invertible disformal transformation, which, nevertheless, introduces branch of solutions with mimetic dark matter. Let us first establish when a disformal transformation is invertible. A general disformal transformation is given as
\begin{equation}
    g_{\mu\nu}=C(Y,\phi)h_{\mu\nu}+D(Y,\phi)\partial_{\mu}\phi\,\partial_{\nu}\phi\ ,\label{eq: mim: disformal transformation}
\end{equation}
where we have denoted the kinetic term with respect to $h_{\mu\nu}$ as $Y$
\begin{equation}
    Y=h^{\mu\nu}\,\partial_{\mu}\phi\,\partial_{\nu}\phi\ .\label{Y}
\end{equation}
The functions $C$ and $D$ are mostly arbitrary; however, they are usually subjected to several conditions, which ensure that the resulting tensor $g_{\mu\nu}$ behaves as a proper metric tensor. In particular we will require $C>0$ and $C+DY>0$, which guarantees that $g_{\mu\nu}$ has a well defined inverse
\begin{equation}
    g^{\mu\nu}=\frac{1}{C}\left [h^{\mu\nu}-\frac{D}{C+DY}\phi^{,\mu}\phi^{,\nu}\right]\ .\label{eq: mim: inverse metric}
\end{equation}
and a non-vanishing determinant \cite{Bekenstein:1992pj,Bettoni:2013diz,Bruneton:2006gf}. In the above expression we have used the notation $\phi^{,\mu}=h^{\mu\nu}\partial_{\nu}\phi$. Note that \eqref{eq: mim: disformal transformation} is at most linear in all but one component of the metric $h_{\mu\nu}$. The only non-triviality stems from the possible complicated dependence on the kinetic term $Y$. Therefore we can invert the relation \eqref{eq: mim: disformal transformation} for nearly all components of $h_{\mu\nu}$ by directly expressing them as
\begin{equation}
    h_{\mu\nu}=\frac{g_{\mu\nu}-D(Y,\phi)\partial_{\mu}\phi\,\partial_{\nu}\phi}{C(Y,\phi)}\ .\label{eq: mim: almost inverse}
\end{equation}
Clearly this is not a true inverse since there is an additional dependence on $Y$ on the right hand side, which still depends on $h_{\mu\nu}$. However, from this form, it is clear that if we are able to express $Y$ through $g_{\mu\nu}$ and $\phi$, then plugging such expression into \eqref{eq: mim: almost inverse} will yield the inverse map to \eqref{eq: mim: disformal transformation}. Fortunately, this can be easily achieved as $Y$ can be related to the kinetic term $X$ defined as  
\begin{equation}
    X=g^{\mu\nu}\,\partial_{\mu}\phi\,\partial_{\nu}\phi\ .\label{X}
\end{equation}
Using expression \eqref{eq: mim: inverse metric} we directly obtain
\begin{equation}
    X(Y,\phi)=\frac{Y}{C+DY}\ .\label{eq: mim: X of Y}
\end{equation}
Hence, if this relation is invertible, then the entire transformation \eqref{eq: mim: disformal transformation} is invertible as well.\par
In \cite{Jirousek:2022rym} we have shown that the crucial condition for appearance of mimetic dark matter is the existence of real solutions for the condition
\begin{equation}
    C=C_{Y}Y+D_{Y}Y^{2}\ ,\label{eq: mimetic singularity condition}
\end{equation}
see \cite{Zumalacarregui:2013pma,Deruelle:2014zza,Jirousek:2022rym,Golovnev:2022jts}. Assuming existence of such solutions has a direct impact on the possible relations between $X$ and $Y$. Indeed, taking a partial derivative of $X$ with respect to $Y$ we find
\begin{equation}
    \partial_{Y}X=\frac{1}{(C+DY)^{2}}\left (C-C_{Y}Y-D_{Y}Y^{2}\right )\ .
\end{equation}
Hence we know that the $X(Y,\phi)$ has a point with a vanishing derivative in the $Y$ direction, whenever the theory supports a mimetic solution. Clearly, if this point is a local maximum or minimum the local invertibility of \eqref{eq: mim: X of Y} is automatically spoiled at the vicinity of the extremum. Consequently, the global invertibility is ruined as well. This clearly leaves us with a single possibility: if expression \eqref{eq: mim: X of Y} is invertible then $C=C_{Y}Y+D_{Y}Y^{2}$ must correspond to an inflection point in the function $X(Y,\phi)$.

\begin{figure}[ht]
\centering 
\includegraphics[scale=0.45]{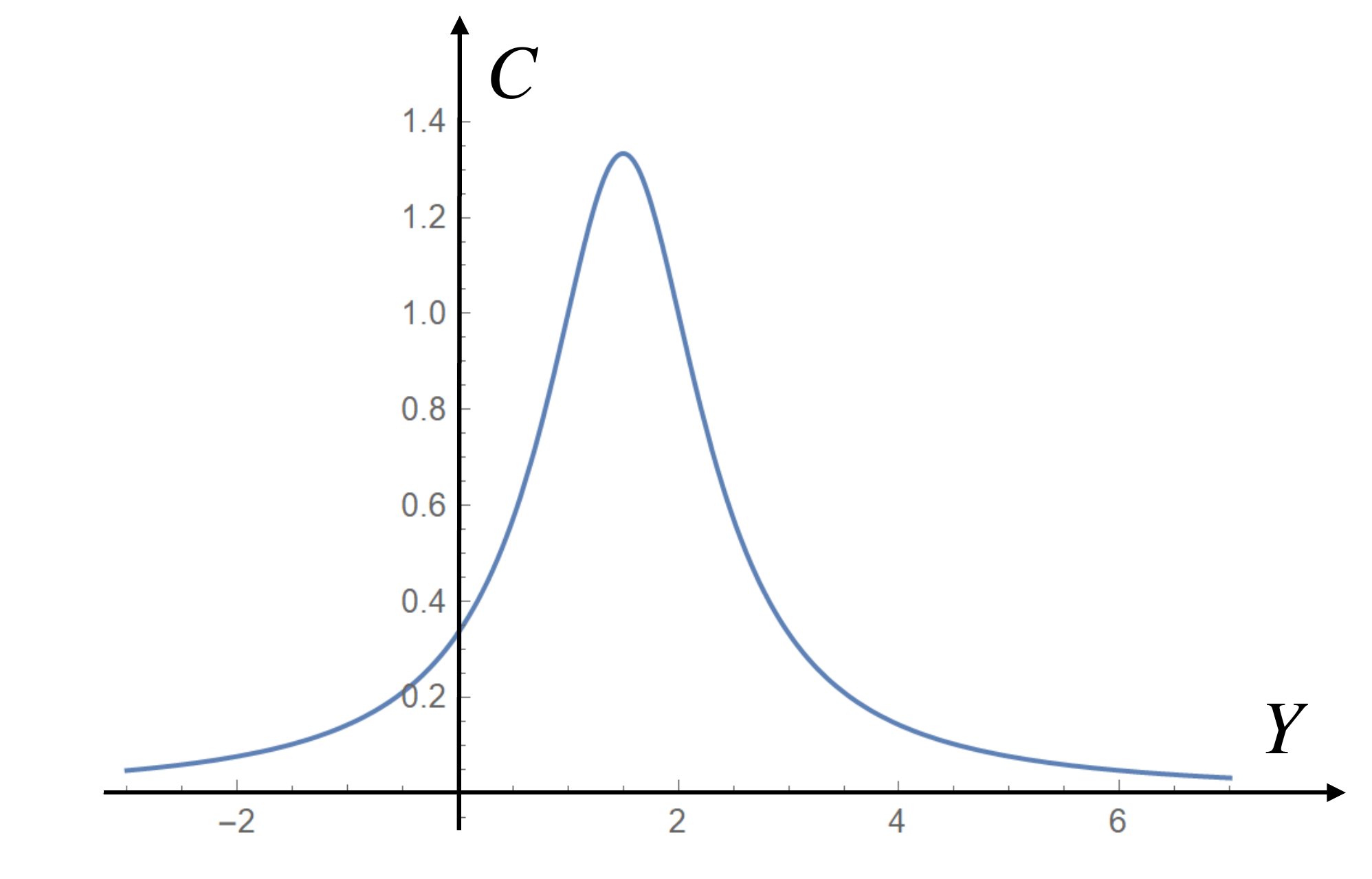}
\caption{\label{fig:1} Graph of the conformal factor $C(Y)$  given by equation \eqref{first_C}. The maximum lies at $Y=3/2$.}
\end{figure}

\begin{figure}[ht]
\centering 
\includegraphics[scale=0.45]{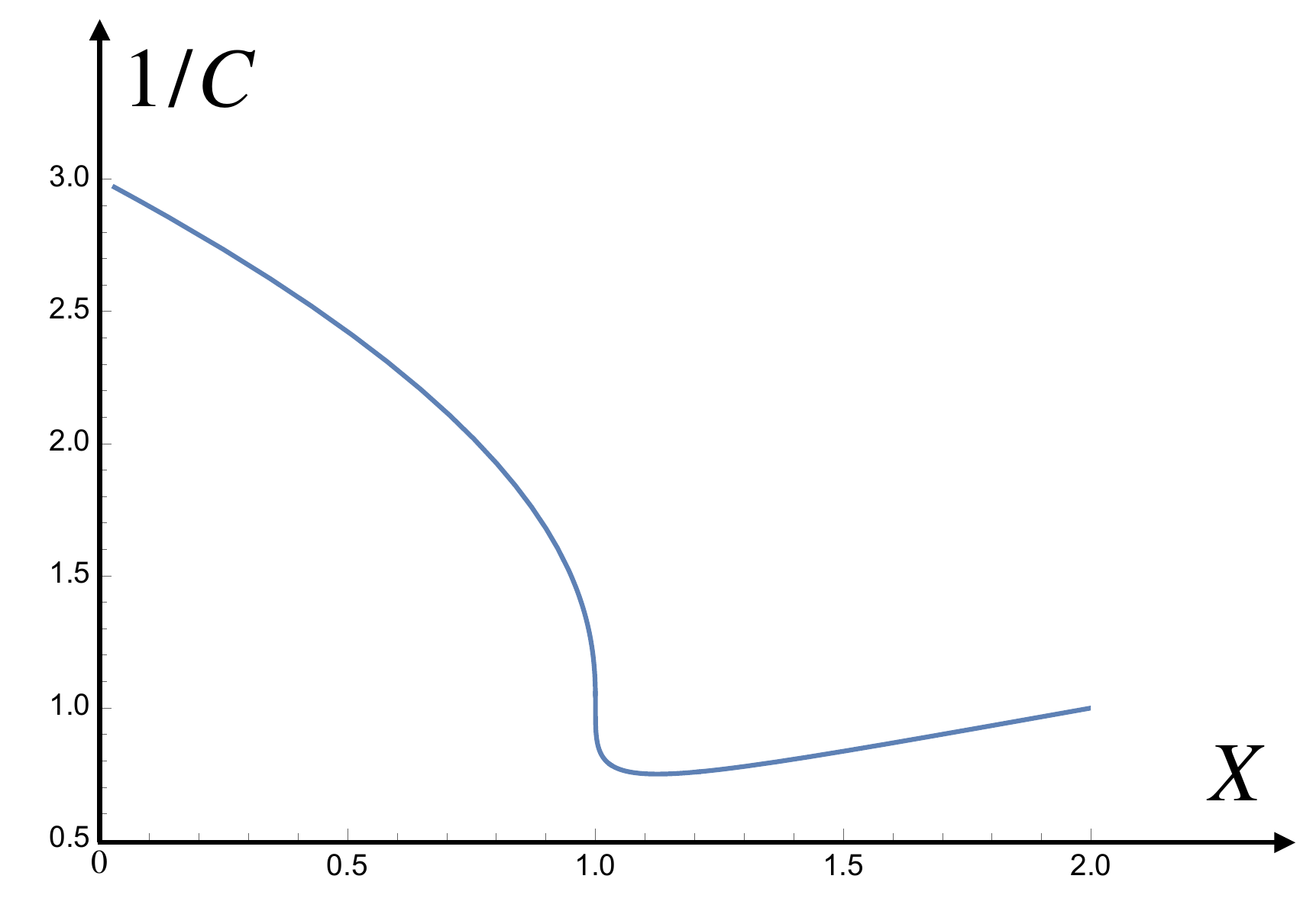}
\caption{\label{fig:2} Graph of the inverse conformal factor  $C^{-1}(Y(X))$ given by equation \eqref{C_inv} as a function of transformed kinetic term $X$ defined in \eqref{X}.
}
\end{figure}

Let us now apply what we have learned to construct a particular transformation that satisfies our requirements. The best place to start is to directly propose the relation between $X$ and $Y$ such that it is invertible and has an inflection point. The simplest such function is the cubic function. Hence we propose
\begin{equation}
    X=(Y-1)^{3}+1\ ,\label{eq: mim: example X of Y}
\end{equation}
with no direct dependence on $\phi$. Note that we offset the inflection point away from $0$ to $Y=1$ because we know that the vanishing of the derivative corresponds to the mimetic solution and a mimetic solution at $Y=0$ implies $C=0$, due to relation \eqref{eq: mimetic singularity condition}. Similarly, we know that the sign of $X$ and $Y$ have to agree due to \eqref{eq: mim: X of Y}. Thus we also offset the inflection point away from $X=0$ to $X=1$. Finally, we set $D=0$ for simplicity. Our choices then give us\footnote{Note that a more general expression of this type
\begin{equation}
    X=(Y-a)^{3}+b\ .
\end{equation}
suffers from several problems whenever $a\neq b$. The resulting functions $C$ will in general have point where it tends to infinity and a point where it passes through zero. At such points the metric metric $g_{\mu\nu}$ might become ill-defined or it is not a metric tensor. No such problems appear for $a=b\neq0$, which is the case here.}.
\begin{equation}
    X=\frac{Y}{C}=(Y-1)^{3}+1\ ,
\end{equation}
which implies
\begin{equation}
    C(Y,\phi)=\frac{Y}{(Y-1)^{3}+1}\ .    \label{first_C}
\end{equation}
Note that we define the value of $C$ at $Y=0$ by its limit which gives $C(0)=1/3$. This function satisfies $C> 0$, see \cref{fig:1}. The second condition $C+DY>0$ is a trivial consequence.
\\
For the conformal factor \eqref{first_C} the metric transformation reads
\begin{equation}
    g_{\mu\nu}=\frac{Y}{(Y-1)^{3}+1}\,h_{\mu\nu}\ .
\end{equation}
The singularity condition \eqref{eq: mimetic singularity condition} gives
\begin{equation}
    C-C_{Y}Y=-\frac{3Y^{2}}{((Y-1)^{3}+1)^2}\left (Y-1\right )^{2}=0\ .
\end{equation}
This equation has a single solution $Y=1$ as intended. Note that for $Y\rightarrow 0$ the above expression has a non-vanishing limit and thus $Y=0$ is not a solution. Therefore we have a single mimetic solution due to this transformation. The expression \eqref{eq: mim: example X of Y} is by design invertible and this inverse can be easily calculated as
\begin{equation}
    Y=\sqrt[3]{X-1}+1\ .
\end{equation}
Plugging this into \eqref{eq: mim: almost inverse} we find the inverse transformation
\begin{equation}
    h_{\mu\nu}=\frac{X}{\sqrt[3]{X-1}+1}\,g_{\mu\nu}\ ,
\end{equation}
which clearly exists. The conformal factor in the above expression is obtained as
\begin{equation}
    C^{-1}(Y(X))=\frac{X}{\sqrt[3]{X-1}+1}\ .
    \label{C_inv}
\end{equation}
As we may expect the inverse transformation is going to have some special features at the singular points of the Jacobian. This can be glanced by differentiating the above conformal factor, which gives
\begin{equation}
    \partial_{X}C^{-1}(Y(X))=-C^{-2}C_{Y}\partial_{X}Y\ .
\end{equation}
We can see that the derivative $\partial_{X}Y$ will tend to infinity since the derivative of the inverse function tends to zero for $Y=1$ or $X=1$. Consequently, the derivative of the conformal factor \eqref{C_inv} will tend to infinity as well, see \cref{fig:2}.
\subsection{Mimetic constraint}
As we have pointed out in \cref{sec:toy}, transformations of the form \eqref{eq: mim: disformal transformation} can be implemented using Lagrange multipliers. For disformal transformations it is convenient to enforce the relation for the inverse metric \eqref{eq: mim: inverse metric} rather then for the metric \eqref{eq: mim: disformal transformation} itself\footnote{It is interesting to note that for standard mimetic gravity from \cite{Chamseddine:2013kea} one can write the constraint in the form 
\begin{equation}
\lambda_{\mu\nu}\left(h^{\alpha\beta}\,\partial_{\alpha}\phi\partial_{\beta}\phi\,\,g^{\mu\nu}-h^{\mu\nu}\right)\,,
\end{equation}
so that the contravariant auxiliary metric $h^{\mu\nu}$ enters linearly as the second Lagrange multiplier on top of the $\lambda_{\mu\nu}$. Here we redefined $\lambda_{\mu\nu}$ by dividing it with $Y$.  
}:
\begin{equation}
    \lambda_{\mu\nu}\left [g^{\mu\nu}-\frac{1}{C}\left ( h^{\mu\nu}-\frac{D}{C+DY}\phi^{,\mu}\phi^{,\nu}\right )\right ]\ .\label{eq: mim: full disformal constraint}
\end{equation}
This is because the right hand side of the relation \eqref{eq: mim: inverse metric} depends purely on $h^{\mu\nu}$, rather then on the mixture of $h^{\mu\nu}$ and $h_{\mu\nu}$ as in \eqref{eq: mim: disformal transformation}. In our previous paper \cite{Jirousek:2022rym} we have shown that the entire effect of a disformal transformation can be captured by a single constraint. Therefore it is clear, that there must be a great deal of redundancy in the above expression. It should be possible to reduce the set \eqref{eq: mim: full disformal constraint} to this single vital constraint.\footnote{A similar construction has been suggested for the special case of mimetic dark matter in \cite{Golovnev:2013jxa} without providing much detail.} Here we show concrete steps how this can be achieved by reducing the full expression \eqref{eq: mim: full disformal constraint} into the constraint found in \cite{Jirousek:2022rym}. We begin by decomposing the Lagrange multipliers $\lambda_{\mu\nu}$ as
\begin{equation}
    \lambda_{\mu\nu}\equiv\Bar{\lambda}_{\mu\nu}+\lambda \partial_{\mu}\phi\,\partial_{\nu}\phi\ .\label{eq: mim: lambda decomposition}
\end{equation}
The barred $\Bar{\lambda}_{\mu\nu}$ is defined to be orthogonal to $\phi^{,\mu}\phi^{,\nu}$. That is
\begin{equation}
    \Bar{\lambda}_{\mu\nu}\phi^{,\mu}\phi^{,\nu}\equiv0\ .\label{eq: mim: lambda orthogonality}
\end{equation}
Clearly the above decomposition is not a good reparametrization of the tensor $\lambda_{\mu\nu}$ when $Y=0$. For such cases the entire right hand side of \eqref{eq: mim: lambda decomposition} vanishes after contraction with $\phi^{,\mu}\phi^{,\nu}$, which implies $\lambda_{\mu\nu}\phi^{,\mu}\phi^{,\nu}=0$. This is clearly not the case for a generic value of $\lambda_{\mu\nu}$. Hence, the case for $Y=0$ must be analysed separately. These cases are of less interest to us since any configuration with $Y=0$ cannot yield a well behaved mimetic solution since the condition for mimetic solutions \eqref{eq: mimetic singularity condition} then implies $C=0$. This leads to the inverse metric $g^{\mu\nu}$ being ill-defined. Proceeding with $Y\neq 0$ allows us to rewrite constraint \eqref{eq: mim: full disformal constraint} as
\begin{equation}
    \Bar{\lambda}_{\mu\nu}\left (g^{\mu\nu}-\frac{1}{C}h^{\mu\nu}\right )+\lambda \left (X-\frac{Y}{C+DY}\right)\ .
\end{equation}
It is important to note, that the decomposition \eqref{eq: mim: lambda decomposition} is no different from a field redefinition of old variables $\lambda_{\mu\nu}$ by novel variables $\bar{\lambda}_{\mu\nu}$ and $\lambda$. Therefore, we must be wary of any possible singular points of the Jacobian of such transformation. However, it turns out that the Jacobian here is regular and finite for $Y\neq 0$. Hence, the field redefinition \eqref{eq: mim: lambda decomposition} is well behaved and does not introduce any additional dynamical content to the theory. In the limit $Y\rightarrow0$ we encounter terms of the Jacobian matrix which tend to infinity, which reflects the ill-defined nature of the transformation for $Y=0$.\par
Similarly to \eqref{eq: mim: lambda decomposition} we decompose the metric $h^{\mu\nu}$ in the following way
\begin{equation}
    h^{\mu\nu}\equiv\Bar{h}^{\mu\nu}+\theta g^{\mu\nu}\ ,\label{eq: mim: h decomposition}
\end{equation}
where we define $\Bar{h}^{\mu\nu}$ to be orthogonal to $\partial_{\mu}\phi\,\partial_{\nu}\phi$
\begin{equation}
    \Bar{h}^{\mu\nu}\partial_{\mu}\phi\,\partial_{\nu}\phi\equiv0\ .\label{eq: mim: h orthogonality}
\end{equation}
The scalar field $\theta$ gives a direct relation between the kinetic term $Y$ and $X$. Indeed, by contracting \eqref{eq: mim: h decomposition} with $\partial_{\mu}\phi\,\partial_{\nu}\phi$ we obtain
\begin{equation}
    Y=\theta\,X\ .\label{eq: mim: Y theta X}
\end{equation}
Again the field redefinition \eqref{eq: mim: lambda decomposition} is only consistent for $X\neq0$, which we shall assume for now. With this choice the decomposition \eqref{eq: mim: h decomposition} has a regular and finite Jacobian. Applying \eqref{eq: mim: h decomposition} and \eqref{eq: mim: Y theta X} allows us to rewrite the constraint as
\begin{equation}
    \Bar{\lambda}_{\mu\nu}\left (g^{\mu\nu}-\frac{1}{C(\theta X,\phi)}(\bar{h}^{\mu\nu}+\theta g^{\mu\nu})\right )+\lambda \left (X-\frac{\theta X}{C(\theta X,\phi)+D(\theta X,\phi)\,\theta X}\right)\ .
\end{equation}
We first take a closer look at the first of the two constraints. In particular we notice that the field $\bar{h}^{\mu\nu}$ now enters strictly linearly. In other words it acts as a Lagrange multiplier, which enforces
\begin{equation}
    \Bar{\lambda}_{\mu\nu}=f_{1}\, \partial_{\mu}\phi\,\partial_{\nu}\phi\ .
\end{equation}
Here $f_{1}$ is a free function, which appears because the variation of $\bar{h}^{\mu\nu}$ is restricted by the condition \eqref{eq: mim: h orthogonality}.\footnote{For a more detailed discussion of restricted variations please see \cite{Jirousek:2020vhy}} However, $f_{1}$ can be immediately determined using the condition \eqref{eq: mim: lambda orthogonality}. Clearly as long as $Y\neq 0$, $f_{1}$ is forced to vanish. In a similar fashion the variation with respect to the Lagrange multiplier $\bar{\lambda}_{\mu\nu}$ implies
\begin{equation}
    \bar{h}^{\mu\nu}=(C-\theta)g^{\mu\nu}+f_{2}\,\phi^{,\mu}\phi^{,\nu}\ .
\end{equation}
The free function $f_{2}$ appears due to the condition \eqref{eq: mim: lambda orthogonality}. Conversely the condition \eqref{eq: mim: h orthogonality} allows us to determine $f_{2}=(\theta-C)X/Y^{2}$. Since the constraint associated with $\bar{\lambda}_{\mu\nu}$ can be always satisfied and $\bar{\lambda}_{\mu\nu}$ is forced to vanish on shell it follows that this constraint does not affect the other equations of motion in any way. Consequently we can integrate it out of the action without affecting the theory. Hence, we are left with a single constraint
\begin{equation}
    \lambda \left (X-\frac{\theta X}{C(\theta X,\phi)+D(\theta X,\phi)\,\theta X}\right)\ .
\end{equation}
Note that since $Y,X,\theta,C$ and $C+DY$ are assumed to be non-vanishing we may redefine the Lagrange multiplier
\begin{equation}
    \bar{\lambda}\equiv\lambda \, \frac{\theta X}{C(\theta X,\phi)+D(\theta X,\phi)\,\theta X}\ ,
\end{equation}
after which the above constraint becomes
\begin{equation}
    \bar{\lambda} \left (\theta^{-1}C(\theta X,\phi)+D(\theta X,\phi)X-1\right)\ .
\end{equation}
This is exactly the constraint found in \cite{Jirousek:2022rym}. Notice that this constraint has a similar structure to \eqref{eq:L_with_Q}. Namely, the auxiliary variable $\theta$ has the same footing as $Q$, since its equation of motion introduces the branching structure of the solutions. Indeed, the equation of motion for $\theta$ gives
\begin{equation}
    \theta^{-2}\bar{\lambda}\Big (C(\theta X,\phi)-C_{Y}(\theta X,\phi)\theta X-D_{Y}(\theta X,\phi)\theta^{2}X^{2}\Big )=0\ .\label{eq: mim: theta EoM with Lambda}
\end{equation}
which clearly consists of two factors, corresponding to the two branches. The regular branch here is also represented by the vanishing of the Lagrange multiplier $\bar{\lambda}$, while the singular branch is characterized by a specific solution for the scalar $\theta$. This solution can be determined algebraically from the second factor of the above equation of motion, that is
\begin{equation}
    C(\theta X,\phi)-C_{Y}(\theta X,\phi)\theta X-D_{Y}(\theta X,\phi)\theta^{2}X^{2}=0\ .\label{eq: mim: theta equation auxiliary}
\end{equation}
This singular branch then contains the mimetic solutions discussed in \cite{Jirousek:2022rym}.
\section{New Form of k-Essence}\label{New_K-essence}
As another example of interesting new degrees of freedom arising from the invertible transformations, let us consider usual general covariant and Lorentz invariant k-essence (see e.g. \cite{Armendariz-Picon:2000nqq,Armendariz-Picon:1999hyi,Chiba:1999ka})
with Lagrangian
\begin{equation}
\mathcal{L}=P\left(X\right)-V\left(\varphi\right)\,,
\label{eq:k-essence}
\end{equation}
 where as before 
\begin{equation}
X=g^{\mu\nu}\partial_{\mu}\varphi\,\partial_{\nu}\varphi\,.\label{eq:X_phi}
\end{equation}
For simplicity, let us restrict our attention to Minkowski spacetime
$\eta_{\mu\nu}$. The corresponding equation of motion is 
\begin{equation}
\frac{\delta S}{\delta\varphi}=-\partial_{\mu}\left(P'\partial^{\mu}\varphi\right)-V'=0\,.\label{eq:phi_EoM_k-essence}
\end{equation}
It is important to note that, for a vanishing potential, (\ref{eq:phi_EoM_k-essence})
corresponds to the conservation of the Noether current 
\begin{equation}
J^{\mu}=P'\partial^{\mu}\varphi\,,\label{eq:current_k-essence}
\end{equation}
 of the shift-symmetry $\varphi\rightarrow\varphi+c$. Now let us
perform an invertible transformation 
\begin{equation}
\varphi\rightarrow\Phi^{3}+\dot{\chi}\,,\label{eq:transformation_k-essence}
\end{equation}
where we explicitly break Lorentz invariance and select a foliation.
This transformation is similar to \eqref{eq: ex1: transformation}. Now the new equations of motion
are 
\begin{equation}
\frac{\delta S}{\delta\Phi}=3\Phi^{2}\,\frac{\delta S}{\delta\varphi}\,,\qquad\text{and}\qquad\frac{\delta S}{\delta\chi}=-\frac{d}{dt}\frac{\delta S}{\delta\varphi}\,.\label{eq:new_EoM_k-essence}
\end{equation}

Thus, as long as $\Phi\neq0$ the equation of motion for $\Phi$ implies
that the equation of motion for $\chi$ is automatically satisfied
so that $\chi$ is purely a gauge degree of freedom. In that case
whole dynamics is indistinguishable from k-essence with original Lagrangian
(\ref{eq:k-essence}). However, when $\Phi=0$ one obtains that $\Phi-$equation
of motion is identically satisfied, while the $\chi-$equation of
motion carries new information, as now 
\begin{equation}
\frac{\delta S}{\delta\varphi}=-q\left(\mathbf{x}\right)\,,\label{eq:introducing_source}
\end{equation}
where $q\left(\mathbf{x}\right)$ is an arbitrary function of spatial
coordinates only. Thus, the original current $J^{\mu}$ is no longer
conserved even in the absence of potential, but produced by a source
$q\left(\mathbf{x}\right)$ as 
\begin{equation}
\partial_{\mu}J^{\mu}+V'=q\left(\mathbf{x}\right)\,.\label{eq:current_q}
\end{equation}
Hence, the dynamics are described by the second order PDE 
\begin{equation}
P^{'}\Box\varphi+2P''\partial^{\alpha}\varphi\,\partial_{\alpha}\partial_{\beta}\varphi\,\partial^{\beta}\varphi+V'=q\left(\mathbf{x}\right)\,,\label{eq:PDE}
\end{equation}
supplemented with a first order equation ODE
\begin{equation}
\dot{\chi}=\varphi\,.\label{eq:def_u}
\end{equation}
These equations of motion can be obtained from a lower derivative
Lagrangian with a constraint 
\begin{equation}
\mathcal{L}\left[q,\varphi,\chi\right]=P\left(X\right)-V\left(\varphi\right)-q\left(\dot{\chi}-\varphi\right)\,,\label{eq:new_Lagrangian}
\end{equation}
with explicitly broken Lorentz symmetry. Comparing with the usual
first order Lagrangian $L=p\dot{q}-H$ we obtain that $\left(-q\right)$
is the canonical momentum conjugated to $\chi$. Importantly, this
momentum is a constant of motion - stays a time-independent function. The total number of DoF is two: $(\phi,\chi)$, as one has to provide four functions on a Cauchy hypersurface. Though, $\chi$ is not a usual wave-like propagating DoF.  
The corresponding Hamiltonian, $H-q\varphi$ emulates the k-essence
(\ref{eq:k-essence}) in the presence of an additional linear potential $\left(-q\left(\mathbf{x}\right)\varphi\right)$
where the coefficient in front depends on initial data. Crucially,
if the potential $V\left(\varphi\right)$ contains at least a mass term
$\frac{1}{2}m^{2}\varphi^{2}$, it will always overcome the dangerous linear
term. Note that the total Hamiltonian of this system is unbounded from bellow due to the freedom in the choice of $q(\mathbf{x})$, which enters the Hamiltonian linearly. However, since $q(\mathbf{x})$ is an integral of motion, the arbitrarily low energies of the total Hamiltonian cannot be accessed dynamically and thus the Ostrogradsky instability cannot be exploited (see e.g. \cite{Woodard:2006nt,Woodard:2015zca}). However, the linear term changes the position
of the minimum of the potential energy and effectively introduces
a symmetry breaking. Thus, one can consider variable $\Phi$ as an
order parameter. For $\Phi\neq0$ the system has usual vacuum $\varphi=0$,
while for $\Phi$ the system has inhomogeneous $q-$vacuum $\varphi\left(\mathbf{x}\right)$
which depends on initial data $q\left(\mathbf{x}\right)$. This is
quite an interesting physical behavior for systems with phase transitions. 

In particular, for a canonical system with $P\left(X\right)=\frac{1}{2}X$
and usual quartic potential 
\begin{equation}
V\left(\varphi\right)=\frac{1}{2}m^{2}\varphi^{2}+\frac{1}{4}\lambda\varphi^{4}\,,
\label{eq:quartic_potential}
\end{equation}
with $\lambda>0$ and $m^{2}>0$ the $q-$vacuum corresponds to minimal value of potential energy
\begin{equation}
\frac{1}{2}\left(\partial_{i}\varphi\right)^{2}+\frac{1}{2}m^{2}\varphi^{2}+\frac{1}{4}\lambda\varphi^{4}-q\left(\mathbf{x}\right)\varphi\,,
\end{equation}
which is given by the solution of the following quasilinear elliptic equation 
\begin{equation}
\left(-\Delta+m^{2}\right)\varphi+\lambda\varphi^{3}=q\left(\mathbf{x}\right)\,.
\end{equation}
Classically one could think of $q\left(\mathbf{x}\right)$ as some external parameter. But in quantum physics $q$ will receive unavoidable quantum fluctuations. 
The appearance of new parts of the potential in k-essence as a result of equations of motion for other DoF resembles what happens in the so-called generalized unimodular gravity \cite{Barvinsky:2017pmm,Barvinsky:2020sxl}.
\\
\\
The transformation \eqref{eq:transformation_k-essence} is not Lorentz-symmetric. However, there is an elegant way to introduce a similar construction in a generally covariant and Lorentz-symmetric fashion. Namely, instead of \eqref{eq:transformation_k-essence}  one writes 
\begin{equation}
\varphi\rightarrow\Phi^{3}+\nabla_{\mu}U^{\mu}\,.
\end{equation}
In this way, the second of equations in \eqref{eq:new_EoM_k-essence} becomes 
\begin{equation}
\frac{\delta S}{\delta U^{\mu}}=-\nabla_{\mu}\frac{\delta S}{\delta\varphi}\,,
\end{equation}
resulting in the source function $q\left(\mathbf{x}\right)$ being space-independent constant $q_0$. In that case the $q$-vacua are Lorentz-symmetric. In particular, for \eqref{eq:quartic_potential} the $q$-vacuum breaks $Z_2$ symmetry $\varphi\rightarrow-\varphi$. Moreover, the excitations around $q$-vacuum have masses different from $m$ and possess cubic self-interaction absent in the original potential \eqref{eq:quartic_potential}.

\section{Conclusions and Discussion}
Following the arguments of our previous work \cite{Jirousek:2022rym}, we have explicitly shown that an invertible transformation can change the number of degrees of freedom. This phenomenon may sound strange and would seemingly contradict the very definition of an invertible transformation. Indeed, by definition, invertibility means a one-to-one correspondence between original and new variables or fields. However, based on the action principle, the dynamics of a system is given by a stationary path with vanishing variation, that is, satisfying the extremum condition. After a transformation of dynamical variables or fields, such an extremum condition can be satisfied not only by the original \emph{regular} stationary path, but also by \emph{singular} configurations, where the Jacobian of the transformation vanishes.\footnote{Just to recall: for a usual function $f(y)$ with extremum at $y_m$, one can make a transformation $y(x)$. The extremum with respect to $x$ is located at configurations where $df/dy\cdot dy/dx=0$. Thus, on top of $y_m$ one gets new configurations for which $dy/dx=0$.} This is a crucial point of our discussion. 

As is well known, a singular transformation is not necessarily non-invertible, while the inverse function theorem says that a regular transformation with non-vanishing Jacobian is at least locally invertible. Therefore, singular but invertible transformations give rise to novel dynamics absent in the original theory. 


We have verified our arguments by explicitly constructing models in point particle systems. We have discussed the dynamics of both of regular and singular branches of solutions separately.

Then, we have confirmed by the Hamiltonian analysis that the number of DoF can change after an invertible transformation in the singular branch. Moreover, even in cases when the number of DoF is preserved, completely different dynamics can emerge in the singular branch. As a further application, we have discussed novel singular but invertible disformal transformations \cite{Bekenstein:1992pj} of the metric. These transformations result in emergence of a new degree of freedom of the same type as in mimetic gravity \cite{Chamseddine:2013kea}. 
\\
\\
After that in section \ref{New_K-essence} we demonstrated that invertible transformations of the scalar field can introduce new degrees of freedom for general k-essence, \cite{Armendariz-Picon:2000nqq,Armendariz-Picon:1999hyi,Chiba:1999ka}). For usual scalar field with potential the new degree of freedom is harmless and does not lead to the Ostrogradsky instability \cite{Woodard:2006nt,Woodard:2015zca}), even though the singular branch the equations of motion is of the fourth order. Instead, this construction allows to consider different Lorentz-breaking vacua and even nontrivial Lorentz-symmetric vacua spontaneously breaking internal symmetries. 
\\
\\
Our argument will be extended furthermore in several directions. One direction is to use this kind of singular but invertible transformation to find new modified gravity theories such as scalar-tensor theories. 
Some of degenerate higher-order scalar-tensor theories (DHOST) theories \cite{Langlois:2015cwa} can be obtained from the Horndeski construction \cite{Horndeski:1974,Deffayet:2011gz,Kobayashi:2011nu} through an invertible \emph{regular} disformal transformation \cite{Zumalacarregui:2013pma}. Recently, more general invertible regular disformal transformations including higher derivatives were constructed \cite{Takahashi:2021ttd}, which were then applied to Horndeski theories to obtain theories beyond DHOST theories \cite{Takahashi:2022mew}. It might be interesting to apply these kinds of construction to a singular but invertible transformation. In the standard regular case, both gravity theories are physically equivalent without essentially different matter couplings (that is, if we transform the matter couplings as well). But, in our singular case, a transformed theory could represent a completely new theory, even if the number of degrees of freedom is preserved, as in our example from subsection \ref{Non-trivial kin.term}. In general we would like to raise awareness that singular transformations, even invertible once, represent a dangerous game in this regard.
\\
Another direction is to extend our argument to more generic transformations with derivatives. In this case, the standard inverse function theorem does not apply as it is, rather, degeneracy conditions are necessary to prevent the appearance of additional degrees of freedom, see  \cite{Babichev:2019twf,Babichev:2021bim}. However, even in this case, at the final step to judge the invertibility, the same argument of singular nature can appear. It would be interesting to investigate whether such a more generic higher derivative transformation allows for a singular but invertible transformation. If it is possible, such transformations could open another new class of theories after the transformation. Yet another direction is to apply our argument to singular, but invertible transformations in phase space in Hamiltonian formalism. This would constitute an interesting extension of the usual canonical transformations.

To conclude, we think that singular invertible transformations provide an important tool and a playground to construct and analyze variety of physical theories relevant to modified gravity and beyond.

\section*{Acknowledgments}
Our collaboration is supported by the Bilateral Czech-Japanese Mobility Plus Project JSPS-21-12 (JPJSBP120212502). P.~J. acknowledges funding from the South African Research Chairs Initiative of the Department of Science and Technology and the National Research Foundation of South Africa. A.~V. acknowledges support from the Grant Agency of the Czech Republic (GAČR grant 20-28525S). M.~Y. acknowledges financial support from JSPS Grant-in-Aid for Scientific Research No. JP18K18764, JP21H01080, JP21H00069. \\

\bibliographystyle{utphys}
\bibliography{references} 

\appendix
\section{Transformation with higher derivatives}
In this part of the section, we will extend the transformation introduced in the paper to higher derivatives. Namely, we shall consider the transformation of the form,
\begin{equation}
    q = Q^3 +\dot \phi^{(n)}.
\end{equation}

The Hamiltonian analysis of the theory will be done in a similar fashion to the main part of the paper. Using $\theta_n$, rather than $\theta$, the higher derivative Lagrangian reduces to the following first derivative theory,
\begin{equation}
    L\rightarrow\frac{1}{2}\dot{q}(Q,\theta_n)^{2}+\sum_{k=1}^{n-1}\lambda_k(\dot \theta_{n-k}-\theta_{n-k+1})+\lambda_n(\dot \phi -\theta_1)\ .
\end{equation}
The variables of this theory is $(Q,\phi,\theta_k,\lambda_k)$ with $k$ running from $1$ to $n$, thus the phase space has the dimension of $4n+4$. 

The canonical momenta are,
\begin{align}
    p_{Q}=&3Q^2 {\dot{q}}\ ,\\
    p_{{\theta_n}}=& {\dot{q}} ,\\
    p_{\theta_k}=&\lambda_{n-k}\ (1\leq k\leq n-1)\ ,\\
    p_{\phi}=&\lambda_n\ ,\\
    p_{\lambda_k}=&0\ (1\leq k\leq n)\ .
\end{align}
Which leads to (2n+1) primary constraints,
\begin{align}
    p_{\theta_k}=&\lambda_{n-k}\ (1\leq k\leq n-1)\ ,\\
    p_{\phi}=&\lambda_n\ ,\\
    p_{\lambda_k}=&0\ (1\leq k\leq n)\ ,\\
    p_{Q}=&3Q^{2}p_{\theta_n}\ .\label{eq: app: primary constraint 2-high} 
\end{align}
The Hamiltonian will then become,
\begin{align}
    H_{D}=&\frac{1}{2}\left (p_{Q}+p_{\theta_n}\right)^{2}+\sum_{k=1}^{n}\lambda_k\theta_{n-k+1}+\alpha\left (p_{Q}-3Q^{2}p_{\theta_n}\right)\ \nonumber\\
    &\, +\sum_{k=1}^{n}\beta_kp_{\lambda_k}\, +\sum_{k=1}^{n-1}\gamma_k\,(p_{\theta_{n-k}}-\lambda_{k}) +\gamma_n(p_\phi-\lambda_n)\ .
\end{align}
The consistency conditions for the constraints $p_{\lambda_k}=0$, $p_{\theta_k}=\lambda_{n-k}$ and $p_{\phi}- \lambda_n$ computes $\beta_k=0$ and $\gamma_k=\theta_{n-k+1}$. So the Hamiltonian becomes
\begin{equation}
    H_{D}=\frac{1}{2}\left (p_{Q}+p_{\theta_n}\right)^{2}+\sum_{k=1}^{n-1}p_{\theta_{n-k}}\theta_{n-k+1} +p_\phi \theta_1+\alpha\left (p_{Q}-3Q^{2}p_{\theta_n}\right)\ .
\end{equation}
Finally, the consistency condition for primary constraint \eqref{eq: app: primary constraint 2-high} computes
\begin{equation}
   \{p_{Q}-3Q^2p_{\theta_n},H_{D}\}=3 Q^2p_{\theta_{n-1}}=0\ .
\end{equation}
We again find two branches, one being $Q=0$ and the other $p_{\theta_{n-1}}=0$.As for the former branch, it's consistency condition leads to
\begin{equation}
     \{Q,H_D\}=\alpha =0\ .
\end{equation}
All constraints that are derived are second-class. Therefore, since we started off with an $(4n+4)$ dimensional phase space, the DoF count is $(4n+4-2n-2)/2 = n+1$, and we have obtained $n$ extra degrees of freedom compared to the original action.

If we choose the branch $p_{\theta_{n-1}}=0$, however, the consistency condition computes,
\begin{equation}
     \{p_{\theta_{n-1}},H_D\}=p_{\theta_{n-2}}\ .
\end{equation}
This leads to a series of new constraints, $p_{\theta_{n-k}} =0 $ for $ 1\leq k \leq n-1$ and $p_\phi =0$, which are all first class constraints. Noting that \eqref{eq: app: primary constraint 2-high} is also a first-class constraint, the $(4n+4)$ dimensional phase space, is reduced to $(4n+4-2n-2\times(n+1))/2 = 1$, and there are no new degrees of freedom compared to the original action.
\end{document}